 \documentclass[apj,numberedappendix]{emulateapj}

 \usepackage{color}
\usepackage{amsmath}
\usepackage{enumerate}
\usepackage{natbib}
\usepackage{hyperref}
\usepackage{ulem}

\def\simlt{\lower.5ex\hbox{$\; \buildrel < \over \sim \;$}}
\def\simgt{\lower.5ex\hbox{$\; \buildrel > \over \sim \;$}}

\def\beq{\begin{equation}}
\def\eeq{\end{equation}}
\def\ba{\begin{eqnarray}}
\def\ea{\end{eqnarray}}
\def\bB{{\,\mathbf B}}

\def\bj{{\,\mathbf j}}

\def\Sect{{\rm Section}} 
\def\Sects{{\rm Sections}} 

\def\Eq{Equation}
\def\Eqs{Equations}

\def\spl{s_{\rm pl}}
\def\dspl{\dot{s}_{\rm pl}}

\def\Uth{U_{\rm th}}
\def\Uel{U_{\rm el}}
\def\Umag{U_{\rm mag}}
\def\Umelt{U_{\rm melt}}
\def\sigB{\mu_B}
\def\sig{\sigma}
\def\sigcr{\sigma_{\rm cr}}

\def\bmelt{b_{\rm melt}}

\def\bext{b_{\rm ext}}
\def\tohm{t_{\rm ohm}}
\def\dpsiohm{\dot{\psi}_{\rm ohm}}

\def\Ftw{F_{\rm tw}}
\def\Fdiss{F_{\rm diss}}

\newbox\grsign \setbox\grsign=\hbox{$>$} \newdimen\grdimen \grdimen=\ht\grsign
\newbox\simlessbox \newbox\simgreatbox \newbox\simpropbox
\setbox\simgreatbox=\hbox{\raise.5ex\hbox{$>$}\llap
     {\lower.5ex\hbox{$\sim$}}}\ht1=\grdimen\dp1=0pt
\setbox\simlessbox=\hbox{\raise.5ex\hbox{$<$}\llap
     {\lower.5ex\hbox{$\sim$}}}\ht2=\grdimen\dp2=0pt
\setbox\simpropbox=\hbox{\raise.5ex\hbox{$\propto$}\llap
     {\lower.5ex\hbox{$\sim$}}}\ht2=\grdimen\dp2=0pt
\def\simgt{\mathrel{\copy\simgreatbox}}
\def\simlt{\mathrel{\copy\simlessbox}}

\begin{document}

\title{Thermoplastic waves in magnetars}

\author{Andrei M. Beloborodov$^1$ and {Yuri Levin}$^2$}
\affil{$^1$Physics Department and Columbia Astrophysics Laboratory,
Columbia University, 538  West 120th Street New York, NY 10027;
amb@phys.columbia.edu}
\affil{$^2$ Monash Center for Astrophysics and School of Physics, Monash University, Clayton, VIC 3800,
Australia; yuri.levin@monash.edu.au}

\begin{abstract}
Magnetar activity is generated by shear motions of the neutron star surface,
which relieve internal magnetic stresses. An analogy with earthquakes and faults
is problematic, as the crust is permeated by strong
magnetic fields, which greatly constrain crustal displacements. 
We describe a new deformation mechanism that is specific to strongly magnetized
neutron stars.
The magnetically stressed crust begins to move because of a thermoplastic instability, 
which launches a wave that shears the crust and burns its magnetic energy.
The propagating wave front resembles the deflagration front in combustion physics. 
We describe the conditions for the instability,
the front structure and velocity, and discuss implications
for observed magnetar activity.
\end{abstract}



\section{Introduction}

The activity of magnetars is explained by their crustal motions that twist the neutron star
magnetosphere (Thompson et al. 2002) and ignite electron-positron discharge  
(Beloborodov \& Thompson 2007). 
The continual $e^\pm$ flow around the magnetar generates the observed hard X-ray
emission (Beloborodov 2013; An et al. 2013; Hasco\"et et al.~2013).
Magnetospheric twisting also explains the variable spindown rate and 
 may be responsible for triggering flares 
 (Thompson \& Duncan 1996; Lyutikov 2006; Parfrey et al. 2013). 

Theoretical calculations show that a magnetospheric twist dissipates in a few years,
in agreement with the timescale of flux decay in transient magnetars 
(e.g. Mereghetti 2008) and observations of slowly shrinking hot spots 
(Gotthelf \& Halpern 2007; Beloborodov 2011). 
The twist injection must occur faster than dissipation, so the twist is created by
crustal motions with vorticity $>1$~rad~yr$^{-1}$.
This is orders of magnitude faster than the accumulation of internal 
stresses due to ambipolar diffusion or
Hall drift (Goldreich \& Reisenegger 1992; Thompson \& Duncan 1996),
and hence  a quickly developing instability is involved in the crust yielding 
to the stresses.

How this instability occurs is an open question. 
Cracks are impossible because the crust has huge pressure (Jones 2003). 
Large slips are forbidden by the presence of strong magnetic fields,  
unless they are aligned within $10^{-3}$~radian
with magnetic flux surfaces (Levin \& Lyutikov 2012).
One plausible yielding mechanism is a plastic flow.
This possibility is especially attractive 
for the inner crust, which may have an alloy-type lattice (Kobyakov \& Pethick 2014).

The internal stresses gradually accumulate
over the lifetime of the star (Vigan\`o et al. 2013). They 
cause elastic crustal deformations until the stress exceeds a critical value $\sigcr$. 
A significant plastic motion can be initiated at this point. 
Its development is assisted by heat generated by the plastic flow. 
We will show below that this leads to an instability launching a thermo-plastic wave
(TPW) resembling a deflagration front. The propagating wave  dissipates 
the magnetic energy inside the crust and creates the external magnetic twist.


\section{Basic equations}

We will describe the mechanism of TPW using the simplest configuration where the 
crust is a slab with density $\rho(z)$ threaded by uniform vertical magnetic field $B_z$. 
The hydrostatic balance between two strong vertical forces (gravity and pressure gradient) 
constrains possible crustal displacements to the horizontal $x$-$y$ plane. The crust 
is almost incompressible, and therefore $B_z$ remains constant.
The model is effectively one-dimensional:
the horizontal components of magnetic field $B_x$, $B_y$ and
displacement $\xi_x$, $\xi_y$ are functions of $z$ only.
Furthermore, for simplicity we take $B_x=0=\xi_x$.
An axisymmetric  generalization will be briefly described in \Sect~5.

The plastic flow velocity is much smaller than the Alfv\'en and shear speeds in the crust.
Therefore the elastic and magnetic forces are nearly balanced,
\beq
   \frac{\partial}{\partial z}\left(\frac{B_y B_z}{4\pi}-\sig\right)=0,
\eeq
where $\sig=\sigma_{zy}$ is the shear stress of the crustal lattice. This gives
\beq
\label{eq:balance}
   b\,\sigB-\sig=\sigB\bext,  
\eeq
where
\beq
    \sigB\equiv\frac{B_z^2}{4\pi},  \qquad b\equiv\frac{B_y}{B_z},
\eeq
and $\bext$ describes the magnetic field above the crust. Below, we demonstrate the TPW 
mechanism with $\bext\approx 0$; the effect of $\bext\neq 0$ is considered in \Sect~5.

If the crust deformation was purely elastic, its strain $s=\partial \xi/\partial z$
would be given by
\begin{equation}
   s_{\rm el}=-\frac{\sig}{\mu},
\label{elastic}
\end{equation}
where $\mu$ is the lattice shear modulus. When $\sig$  
exceeds some critical value $\sigcr$, a plastic shear flow is produced. 
This may occur either due to the activation of motion of microscopic lattice defects 
(e.g. Mason 1960), or due to the nucleation of a multitude of magnetically-constrained
micro-cracks (Levin \& Lyutikov 2012). The total strain has two parts,
\beq
\label{eq:spl}
    s=s_{\rm el}+\spl,
\eeq
where $s_{\rm el}$ is given by  Equation~(\ref{elastic}) and $\spl$ is the plastic
part of the strain. We
will use the Bingham-Norton  model of an elastic perfectly viscoplastic solid 
(see e.g. Irgens 2008). It gives the plastic flow rate in the form,
\begin{equation}
   \dot{s}_{\rm pl}=-\frac{\sig-\sigcr}{\eta}\, \Theta(\sig-\sigcr),
\label{eq:plastic}
\end{equation}
where $\sigma>0$ is assumed, $\Theta(X)$ is the Heaviside function, and $\eta$ is 
the effective dynamic viscosity. The behaviour of $\sigcr$ has been studied in detail
for terrestrial alloys. Two features are common: \newline (1) 
Strain hardening: $\sigcr$ may increase with $|s_{\rm pl}|$
due to the trapping of dislocations at the boundaries between the alloy grains. 
On the other hand, recrystallization can occur ---
the nucleation of new domains that introduce new dislocations. 
Both effects are neglected below.
\newline
(2) Thermal softening: $\sigcr$ decreases with 
temperature, due to the increased mobility of dislocations. 
In terrestrial alloys, this leads to the formation of shear bands -- narrow bands with high 
$\dot{s}_{\rm pl}$. In a magnetically stressed neutron-star crust thermal softening 
leads to a TPW. We model the thermal softening in the linear approximation,
\beq
\label{eq:sigcr}
   \sigcr=\sigma_0-\zeta(\Uth-U_0),
\eeq
where $\Uth$ is the thermal energy density of a current state, $U_0$ and 
$\sigma_0$ are the thermal energy density and critical stress of some initial state, and  
$\zeta$ is a positive constant. Its value may be estimated as follows. 
At melting temperature $T_m\sim 3\times 10^9 \rho_{12}^{1/3} {\rm ~K}$
one expects the transition to fluid behavior, $\sigcr=0$. 
The corresponding thermal energy density is $\Umelt\sim \sigma_0/\zeta$.
The ratio of Coulomb lattice energy $U_{\rm Coul}$ to its thermal energy at the 
melting point $\Umelt$ is $\Gamma\simgt 10^2$.  
This gives an order-of-magnitude estimate for $\zeta$,
\beq
\label{eq:zeta}
    \zeta\sim \frac{\sigma_0}{\Umelt}
            \sim \frac{\sigma_0\Gamma}{U_{\rm Coul}}
            \sim \Gamma s_0\,\frac{\mu}{U_{\rm Coul}}.
\eeq
Here $s_0=|s|_{\max}$ is the maximum elastic strain of the cold crust at which it must 
yield, so $\sigma_0=\mu s_0$. At  low $T_0$, in the absence of mobile dislocations,
one would expect $s_0\sim 0.1$, comparable to the yielding threshold for an ideal 
crystal (Chugunov \& Horowitz 2010). Using $\mu/U_{\rm Coul}\sim 0.1$ (e.g. Chamel 
\& Haensel 2008), one finds $\zeta\simgt 1$.
\Eq~(\ref{eq:sigcr}) can be replaced by more realistic models of thermal 
softening (Chugunov \& Horowitz 2010). The true $\zeta$ is a function of 
temperature $T$; it is sensitive to the behavior of heat capacity $c_V(T)$ and 
depends on neutron superfluidity.

In contrast to an elastic response, the plastic flow is an irreversible dissipative 
process. It is accompanied by the heating rate per unit volume $\dot{q}=-\sig\dspl$.
Heating happens only where the plastic flow occurs, and this can lead to a strong 
temperature gradient $\partial T/\partial z$ and significant heat conduction.
The thermal equation for the crustal material then reads
\beq
\label{eq:th1}
   \frac{\partial \Uth}{\partial t}=-\sigma \dspl + \chi\frac{\partial^2 \Uth}{\partial z^2},
\eeq
where $\chi\approx \kappa/c_V$ the heat diffusion coefficient.
Here the simplest model assumes $\chi=const$;
using the estimates of thermal conductivity $\kappa$ and heat capacity $c_V$ 
(e.g. Gnedin et al. 2001), one finds the characteristic $\chi\sim 10$~cm$^2$~s$^{-1}$ 
in the deep crust. \Eq~(\ref{eq:th1}) neglects
neutrino cooling; it is assumed to be small compared 
with plastic heating and heat conduction on the timescales of interest.

The crust is an excellent conductor, and we consider its dynamics on 
timescales much shorter than the timescale for ohmic dissipation. 
The field is therefore advected by horizontal displacement of the crust so that 
\begin{equation}
  \dot{b}=\dot{s},
  \label{eq:drozen}
\end{equation}
\beq
\label{eq:sb}
    b=s-s_0+b_0,  
\eeq
where $b_0$ and $s_0$ are the values of $b$ and $s$ at the onset of plastic flow.

Equations (\ref{elastic})-(\ref{eq:sigcr}),
(\ref{eq:th1}), and (\ref{eq:sb}) specify the plastic flow dynamics.


\section{Local conversion of magnetic energy to heat}

Suppose $B_y$ is quickly dissipated above the crust, leaving no residual 
external stress. Then
$\bext\approx 0$ and
 \Eq~(\ref{eq:balance}) gives
\beq
\label{eq:balance1}
  \sig=b\sigB.
\eeq
The density of ``free'' energy 
(elastic + magnetic) is given by
\beq
   \Uel+\Umag=\frac{\mu s_{\rm el}^2}{2}+\frac{\sigB b^2}{2},
\eeq
and the stress balance~(\ref{eq:balance1}) implies
\beq
\label{eq:en_ratio}
   \frac{\Uel}{\Umag} =-\frac{s_{\rm el}}{b}=\frac{\sigB}{\mu}. 
\eeq

Using \Eq~(\ref{eq:balance1}), $\spl=s+\sig/\mu$, and $ds=db$, 
one obtains the following 
expression for the heat generated by the plastic flow:
\beq
\label{eq:heat}
   dq=-\sig d\spl =-d\left[\sigB\,\frac{b^2}{2}\left(1+\frac{\sigB}{\mu}\right)\right].
\eeq
It expresses energy conservation: $dq=-d\Umag-d\Uel$.

For example, consider a magnetar with $B_z= 4\times 10^{14}$~G, which 
corresponds to $\sigB=B_z^2/4\pi\approx 10^{28}$~erg~cm$^{-3}$. 
In the deep crust, the shear modulus $\mu\sim 10^{28} \rho_{12}^{4/3}$~erg~cm$^{-3}$
exceeds $\sigB$. In this case, in a magneto-elastic equilibrium 
the stored elastic energy is negligible compared to the 
energy stored in $B_y$: $\Umag\gg\Uel$ and $|b/s_{\rm el}|\gg 1$; 
hence $|s_{\rm el}|\ll |\spl|$ and $\spl\approx s$.

It is useful to first consider the plastic flow neglecting heat 
conduction; this approximation will be relaxed in \Sect~4. Then the 
plastically generated heat remains stored locally in the crust, $d\Uth=dq$, and 
\beq
\label{eq:Uth}
   \Uth-U_0=\sigB\,\frac{b_0^2-b^2}{2}  \qquad   (\chi=0, \;\, \mu\gg\sigB).
\eeq
For definiteness, we
assume $b>0$; then $s<0$ and $\sig>0$.
 \Eqs~(\ref{eq:plastic}), (\ref{eq:sigcr}), and (\ref{eq:Uth}) give
\beq
\label{eq:sdot1}
   -\eta\dspl = \sig-\sigma_{\rm cr} = b\sigB-\sigma_0+\zeta\,\sigB\,
      \frac{b_0^2-b^2}{2}.
\eeq
The plastic flow starts (and stops) at $\sig=\sigcr$; 
$\sig>\sigcr$ during the flow. 
The flow is initiated at $b=b_0$ and shears the crust so that $b$ is reduced.

It is instructive to rewrite \Eq~(\ref{eq:sdot1}) in the following form,
\beq
\label{eq:sdot}
       \eta\dot{b}=\frac{\zeta\,\sigB}{2}\, (b-b_0)(b-b_1),
\eeq
where we used $\dspl\approx\dot{s}=\dot{b}$ and 
\beq
   b_0=\frac{\sigma_0}{\sigB}, \qquad 
   b_1=\frac{2}{\zeta}-b_0.
\eeq
Two conclusions can be drawn from \Eqs~(\ref{eq:sdot1}) and (\ref{eq:sdot}):

\noindent
(1) $d\sigcr/db>0$ due to heating with decreasing $b$, and
a quick plastic relaxation of the magnetic stress is triggered at $b_0$
if $d\sigcr/db>d\sig/db=\sigB$.
Then a plastic deformation reducing $b$ results in $\sig>\sigcr$, which 
continues to drive $b$ away from $b_0$ with an increasing rate;
this rate is controlled by viscosity $\eta$: $\dot{b}\approx -(\sig-\sigcr)/\eta$.
The condition for this instability, $d\sigcr/db>\sigB$, can be re-written as
\beq
\label{eq:inst}
  \frac{d\sigcr}{d\Uth}\,\frac{dq}{db} > \sigB.
\eeq
In our simple model with the linear $\sigcr(\Uth)$ this condition becomes $\zeta\, b_0>1$,
which is equivalent to $b_1<b_0$. If this condition is not satisfied, the stress relaxation 
through the thermoplastic instability does not occur; instead there is a slow creep with 
$\sig=\sigcr$. 
The magneto-elastic balance at the onset of instability gives $b_0=(\mu/\sigB)s_0$
(\Eq~\ref{eq:en_ratio}).
One can see that $b_0\gg s_0$ as long as $\mu\gg\sigB$, and hence the condition 
$\zeta\,b_0>1$ can be easily satisfied in the deep crust. 

\noindent
(2) Two final states are possible for the unstable plastic flow. 
If $b_1>0$, the flow stops when $b$ is reduced to $b_1$; then it freezes with 
$\sig>0$. The total energy density dissipated by the plastic flow is then given by 
\beq
    q=\frac{\sigB}{2}\,\left(b_0^2-b_1^2\right),
    \qquad (b_1>0).
\eeq
If $b_1<0$, $\sigcr$ vanishes at some $\bmelt$ ($0<\bmelt<b_0$).
In this case, the crust melts during the plastic flow and becomes fluid, with negligible 
elastic stress; the magnetic field will then relax to the stress-free state with $b\approx 0$.
Then all free energy is converted to heat, $q\approx\sigB\,b_0^2/2$.

The temperature achievable by converting magnetic energy to heat is 
$T\sim B_y^2/8\pi c_V$. Using the heat capacity 
$c_V\sim 10^{18}-10^{19}$~erg~cm$^{-3}$~K$^{-1}$ (its exact value depends on 
neutron superfluidity, Gnedin et al. 2001), one finds 
$T\sim 10^{9}-10^{10}B_{y,15}^2$~K. 
It may exceed the melting temperature. 

The characteristic timescale of the unstable flow may be estimated from  
\Eq~(\ref{eq:sdot}). Besides time $t$, it contains
only two quantities of non-zero dimension, $\sigB$ and $\eta$.
Their ratio defines the timescale for the local energy release, 
\beq
\label{eq:time}
     \Delta t \sim \frac{\eta}{\sigB}.
\eeq


\section{Thermoplastic wave}

A TPW is analogous to a deflagration front. 
Both phenomena operate on two essential ingredients: 
(1) Local heating/burning rate increases with temperature. 
For deflagration this is a chemical reaction while for TPW this is 
the temperature-sensitive plastic flow.
(2) Heat conduction helps the burning to spread. In our case, the horizontal 
magnetic field plays the role of unburned fuel; it is ``consumed'' by the induced plastic flow.  

Consider a crust that is stressed by a slowly evolving magnetic field so that $\sig$ 
gradually approaches the critical value $\sigcr$.
It first reaches $\sigcr$ at some $z_0$.
The plastic flow initiated at $z_0$ generates heat, which is conducted to the 
neighborhood of $z_0$ where $\sig(z)$ is initially smaller than $\sigcr(z)$.
The spreading heat reduces $\sigcr$ and 
can lead to $\sigcr<\sig$ 
so that the region of plastic flow can spread upward and downward. 
When the characteristic thickness of the heated layer reaches $l$ 
estimated below, heat conduction is marginally able to cool it. 
Then the thermoplastic instability launches a wave resembling 
the deflagration front in combustion physics. 

The propagation speed of the TPW
is determined by the heat diffusion coefficient $\chi$
and viscosity $\eta$. Recall that $\eta$ sets the 
characteristic timescale of ``burning" the magnetic energy density $\Delta t$ 
(\Eq~\ref{eq:time}) 
while the characteristic width of the burning front $l$ is related to $\chi$ by
\beq
  l \sim (\chi\,\Delta t)^{1/2}.
\eeq
The front velocity is 
\beq
\label{eq:v}
       v\sim \frac{l}{\Delta t}\sim \left(\frac{\chi \sigB}{\eta}\right)^{1/2}.
\eeq

The parameters of the crust ahead of the front vary with $z$ on the hydrostatic
scale $H$. The heating front with the characteristic thickness $l\ll H$ is
a quasi-steady propagating structure where all parameters are functions
of $u=z-vt$. The front structure is described by 
the plastic stress \Eq~(\ref{eq:plastic}) and the thermal \Eq~(\ref{eq:th1}). 
Using $\partial/\partial t=-vd/du$ and $\partial/\partial z=d/du$, these equations give
\beq
\label{eq1}
  v\eta \frac{db}{du}=\sigB(b-b_0) + \zeta(\Uth-U_0),
\eeq
\beq
\label{eq2}
    \chi\frac{d\Uth}{du}
    =-v\left[\frac{\sigB}{2}(b^2-b_0^2) + \Uth
  \right],
\eeq
where we used \Eq~(\ref{eq:heat}) for the generated plastic heat. 
We choose $u=0$ where the flow is initiated; $\sig=\sigcr(U_0)$ at this point.
Far ahead of the front the crust is cold ($\Uth\approx 0$) and stable,
$\sig<\sigcr(0)$. The ratio $f=\sigcr(0)/\sigcr(U_0)>1$ determines how much 
preheating (through heat conduction) occurs ahead of plastic flow zone.
 
\Eqs~(\ref{eq1}) and (\ref{eq2}) can be solved as follows. 
We divide them and obtain one equation for $w=\Uth/\sigB$
as a function of $b$,
\beq
\label{eq}
   \frac{dw}{db}=-p\,\frac{[w-(b_0^2-b^2)/2]}{\zeta w - b_0 f+b}, 
   \qquad p\equiv\frac{v^2\eta}{\chi\sigB}.
\eeq
This equation has a critical
point where the numerator and denominator must vanish simultaneously.
As long as $2f<\zeta b_0+(\zeta b_0)^{-1}$, this point is reached at 
\beq
  b_1=\frac{1}{\zeta}\left[1-\sqrt{(\zeta b_0-1)^2-2\zeta b_0 (f-1)}\right].
\eeq
We will assume $b_1>0$. Note that $db/du\rightarrow 0$ and $d\Uth/du\rightarrow 0$
as $b$ approaches $b_1$, and hence $b_1$ represents the asymptotic downstream 
state of the crust behind the front. 

For any given parameter $p$ \Eq~(\ref{eq}) can be integrated from $b=b_0$ 
towards smaller $b$, which gives a solution $w(b)$. 
Only one value of $p$ gives $w(b)$ that satisfies the regularity condition at $b_1$. 
We find the required $p$ and the corresponding solution $w(b)$
numerically, using iterations. This determines the front velocity 
$v=(p\chi\sigB/\eta)^{1/2}$.
Then, using the known $v$ and the obtained relation between $b$ and $w$
we integrate \Eqs~(\ref{eq1}) and (\ref{eq2}). A sample solution is shown in Figure~3. 
The model assumes constant diffusion coefficients $\eta$ and $\chi$; 
extension to models with temperature-dependent $\eta$ and $\chi$
is straightforward.

\begin{figure}[t]
\begin{tabular}{c}
\includegraphics[width=0.43\textwidth]{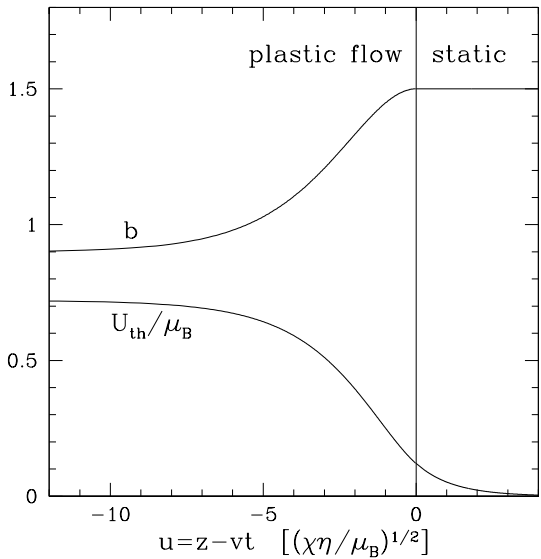} 
\end{tabular}
\caption{
TPW structure calculated for $\zeta=1$, $b_0=1.5$, and $f=1.08$.
In this case, we find $p=0.714$.
}
\label{fig_cartoon}
\end{figure}


\begin{figure}[h]
\begin{tabular}{c}
\includegraphics[width=0.43\textwidth]{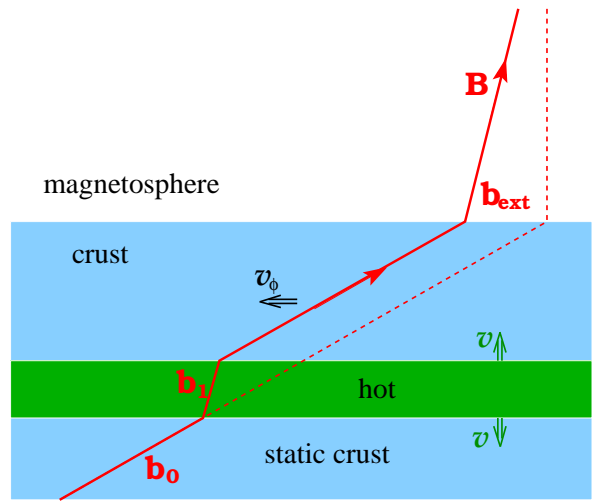} 
\end{tabular}
\caption{
The crustal plate above the plastically heated layer moves with horizontal 
velocity $v_\phi$ that is related to the TPW velocity $v$ 
by the kinematic constraint $v_\phi=-2(b_0-b_1)v$. Magnetic field lines are shown in 
red; they are frozen in the crustal material; $b=B_\phi/B_z$. 
Dotted line shows the magnetic field line before the pair of TPW was launched.
}
\label{fig_cartoon}
\end{figure}


\section{Twisting the external magnetic field}

Consider now an axisymmetric crustal plate in a cylindrical coordinate system 
$(r,\phi,z)$. If gradients along $z$ are much greater than gradients along $r$
(the depth of the crustal plate much smaller than its radius) 
our plate is locally approximated by the slab model
described in \Sects~2-4 with $x=r$, $y=r\phi$, and $B_y=B_\phi$.
Dissipation of toroidal field $B_\phi$ by the thermoplastic front propagating vertically 
inside the plate causes its rotation in the $\phi$ direction (Figure~2). 

Once the pair of upward and downward fronts are launched from the plastically 
heated region, the upper crust starts to rotate with respect to the static lower crust.
Each crustal layer crossed by the front is sheared by $\Delta s= b_1-b_0$,
and propagation of the two fronts in time $dt$ rotates the upper crust by
$d\xi=-2(b_0-b_1)\,vdt$. This gives the rotation velocity of the plate,
\beq
     v_\phi=\frac{d\xi}{dt}=-2(b_0-b_1) v,
\eeq
which twists the external magnetosphere attached to it.
The magnetosphere is force-free, i.e. electric current $\bj$ associated with $B_\phi$
must flow along $\bB$. The magnetospheric current penetrates the plate only where 
rotation is differential in the horizontal plane, $d/dr(v_\phi/r)\neq 0$; 
in the region of rigid rotation the circuit closes horizontally 
along the plate surface (Beloborodov 2009).

Consider a dipole magnetosphere 
and let our plate be a polar cap (or a ring) of radius $r_0=R\sin\theta_0$, where 
$R$ is the star radius. The field component normal to the surface, $B_R$, plays
the role of $B_z$ in previous sections, so $\sigB=B_R^2/4\pi$.
The plate displacement $d\xi$ twists the external 
magnetic field lines by angle $d\psi\approx -d\xi/r_0$ (assuming their 
footprints in the other hemisphere are static). An external stress $\sigB\bext$
is then generated above the plate, where $b=B_\phi/B_R$.
Using the relation between $B_\phi$ and $\psi$ for a moderately twisted dipole 
magnetosphere (Beloborodov 2009), we find
$\bext\approx (\psi/4)(\sin^3\theta_0/\cos^2\theta_0)$.

The rotating plate pumps $\bext$ with rate $\dot{b}_{\rm ext}\propto \dot{\psi}\propto -v_\phi$.
On the other hand,
the external twist is damped on the ohmic timescale $\tohm$, because sustaining 
magnetospheric currents requires voltage $\Phi\sim 10^9-10^{10}$~V
(Beloborodov \& Thompson 2007). The resulting evolution equation for $\bext$ reads
\beq
\label{eq:bext}
     \frac{d\bext}{dt}=\frac{\dot{\psi} \sin^3\theta_0}{4\cos^2\theta_0},\qquad
     \dot{\psi}=\frac{2b_0v}{r_0}-\dot{\psi}_{\rm ohm}.         
\eeq 
A detailed derivation of the ohmic damping 
$\dot{\psi}_{\rm ohm}\sim 1$~rad~yr$^{-1}$ is given in Beloborodov (2011). 
Delivering the external twist $\dot{\psi}>0$ requires a sufficiently fast propagation of 
the thermoplastic front, 
$v\simgt\dpsiohm r_0/2b_0\sim 10^{-3}\, r_{0,\rm km}/b_0$~cm~s$^{-1}$.
Comparing with \Eq~(\ref{eq:v}), we find that this condition translates to 
an upper bound on the (unknown) viscosity of the plastic flow, 
$\eta<\chi\, B_\phi^2/\pi \dpsiohm^2 r_0^2\sim 10^{35}$~erg~s~cm$^{-3}$.

If the rotating plate is near the magnetic pole, $\theta_0<1$, 
even a strong twist $\psi\sim 1$ corresponds to a small $\bext\ll 1$.
In this case, neglecting $\bext$ in the TPW model is a valid assumption.
Significant $\bext$ can be generated at large $\theta_0$.
\Eq~(\ref{eq:balance}) shows how this would reduce the elastic stress inside the plate,
\beq
    \sigma=\sigB(b-\bext).
\eeq
When $\bext$ becomes comparable to $b$, the TPW will be choked.
Later $\bext$ is damped ohmically in the magnetosphere, $\sigma$
rises and the front can be launched again. 

The heating rate in the front is $\dot{q}\sim B_\phi^2/8\pi \Delta t$, where 
$\Delta t$ is related to the front velocity $v$ by $\Delta t\sim \chi/v^2$.
This gives 
\beq
\label{eq:qdot}
    \dot{q}\sim \frac{B_\phi^2\, v^2}{8\pi \chi}.
\eeq
The condition $v>v_{\min}\sim \dpsiohm r_0/2b_0$ corresponds to a minimum heating rate
\beq
   \dot{q}_{\min}\sim 10^{-2}\,\frac{B_R^2\, r_0^2\, \dpsiohm^2}{\chi}.
\eeq
Using $B_R\sim 3\times 10^{14}$~G typical for magnetars, 
$\dpsiohm\sim 1$~rad~yr$^{-1}$, and $\chi\sim 10$~cm$^2$~s$^{-1}$, we estimate
$\dot{q}_{\min}\sim 10^{22}$~erg~cm$^{-3}$~s$^{-1}$. The minimum heating exceeds
neutrino cooling $\dot{q}_{\nu}\sim 10^{21}$~erg~cm$^{-3}$~s$^{-1}$
(e.g. Gnedin et al. 2001), which vindicates
our neglect of neutrino cooling in the TPW model.
Heating is concentrated in the thin front, $l< \chi/v_{\min}\sim 100$~m,
and neutrino emission cools the heated crust {\it behind} the front before it propagates 
a distance comparable to the crust thickness.

Creating the external twist by the TPW is inevitably accompanied by dissipation 
at large depths where heat is drained by neutrino emission (Kaminker et al. 2009).
It is useful to compare the energy flux delivered into the external twist 
$\Ftw=-v_\phi \bext \sigB$ with
the internal dissipation rate per unit area, $\Fdiss\approx v(b_0^2-b_1^2)\sigB$,
\beq
    \frac{\Ftw}{\Fdiss}=\frac{2\bext}{b_0+b_1} 
       \approx \frac{\psi\sin^3\theta_0}{2 (b_0+b_1)\cos^2\theta_0}.
\eeq


\section{Discussion}

The proposed mode of rapid plastic motion 
exists only in materials that are magnetically stressed to  breaking point;
it remains to be seen whether it
could be produced in a high-B terrestrial experiment. TPW is likely to operate in 
magnetars and shape their observed
activity by rapidly dissipating magnetic energy inside the crust 
and by twisting the external magnetosphere.

The thermoplastic front heats the crust so strongly that it may melt it and destroy
neutron superfluidity. The melted material crystalizes behind the front on the 
neutrino cooling timescale. The lifting of superfluidity can transfer angular 
momentum from neutron superfluid to the lattice and produce a spin-up "glitch."

A TPW at low magnetic latitudes is choked when the external $B_\phi$
becomes comparable to the internal 
$B_\phi$, and is revived after the external twist has been ohmically damped.
Repeated crustal shearing is generally expected,
leading to repeating events of magnetar activity. 
The duration of these events is controlled by the speed of the TPW,
which determines the shear rate of the crust. 
This rate cannot be smaller than $\sim 1$~rad~s$^{-1}$ and is likely much 
higher. However, it is not high enough to explain the duration of magnetar 
bursts $\sim 0.1-0.3$~s. The bursts must be triggered in the magnetosphere,
like solar flares, or by a faster failure mode of the crust that is yet to be
identified. The burst
duration is likely set by the mechanism of energy dissipation
taking significantly longer than the Alfv\'en crossing time of the magnetosphere.

Our description of a TPW delivering external twists works for a broad range of 
viscosity $\eta\simlt 10^{35}$~erg~s~cm$^{-3}$.  A weak lower bound 
$\eta>10^{12}$~erg~s~cm$^{-3}$
is provided by electron viscosity (Chugunov \& Yakovlev 2005). A more relevant 
estimate may come from phonon viscosity, as the phonon gas exerts drag on 
the thermally activated dislocations (Mason 1960).

Our future work will extend the TPW model to explicitly include 
the build up of magnetic stresses due to Hall drift. Multidimensional simulations 
could show the beginning and development of the TPW into the 
propagating front described in this Letter. 
Future models will invoke the global crust structure with accurate
heat capacity and thermal conductivity.

\acknowledgements
AMB acknowledges support by NASA grant NNX13AI34G.
YL's research was supported by the Monash Reasearch Acceleration grant. 
We thank Andrei Chugunov for discussions, and Sarah Levin for help with the prose.



\end{document}